# Unveiling Trail Making Test: Visual and manual trajectories indexing multiple executive processes

Linari, Ignacio[1]*; Juantorena, Gustavo E[1]*; Ibañez, Agustín[2]; Petroni, Agustín[1,3]^; Kamienkowski, Juan E[1,4]^

[1] Laboratorio de Inteligencia Artificial Aplicada, Instituto de Ciencias de la Computación, Facultad de Ciencias Exactas y Naturales, Universidad de Buenos Aires - CONICET, Argentina

[2] Cognitive Neuroscience Center (CNC), Universidad de San Andrés, and National Scientific and Technical Research Council (CONICET), Buenos Aires, Argentina; Global Brain Health Institute (GBHI), University of California San Francisco (UCSF), San Francisco, US; and Trinity College Dublin (TCD), Ireland; Latin American Brain Health Institute (BrainLat), Universidad Adolfo Ibáñez, Santiago, Chile.

[3] University of Gothenburg, Sweden

[4] Maestría de Explotación de Datos y Descubrimiento del Conocimiento, Facultad de Ciencias Exactas y Naturales, Universidad de Buenos Aires, Argentina

*,^ Equal contributions

## 1. Abstract

The Trail Making Test (TMT) is one of the most popular neuropsychological tests in the clinical assessment of executive functions (EF) and research in a wide range of clinical conditions. In addition to its sensitivity to executive dysfunction, the TMT presents several strengths: it is simple and intuitive, it is easy to understand for patients, and has a short administration. However, it has important limitations. First, the underlying EFs articulated during the task are not well discriminated, which makes it a test with a low specificity. Second, the traditional pen-and-paper version presents one trial per condition which introduces high variability. Third, only the total time is quantified, which does not allow for a detailed analysis. Fourth, it has a fixed spatial configuration per condition.

In the present study we designed a computerized version of the TMT (cTMT) to overcome its main limitations. Eye and hand positions are simultaneously measured with high spatio-temporal resolution, several trials are acquired, and spatial configuration of the targets is controlled. Our results showed a very similar performance profile compared to the traditional TMT. Moreover, it revealed similarities and differences in eye movements between the two parts of the task. Most importantly, we found an internal working memory measure of the cTMT based on hand and eye movements that showed an association to a validated working memory task. Additionally, we found another internal measure of the TMT, also based on hand and eye movements, that we propose as a potential marker of inhibitory control. Our results showed that executive functions can be studied in more detail using traditional tests combined with powerful digital setups. Finally, our study paved the way for a detailed analysis of other complex tasks used for clinical evaluation, providing a deeper understanding of the processes underlying its resolution.

## 2.   Introduction

The Trail Making Test (TMT) is perhaps the most popular neuropsychological task used for standard clinical assessment and research (Lange et al., 2005; Bowie & Harvey, 2006; Rabin et al., 2007; Reitan, 1958; Salthouse, 2011; Soukup et al, 1998). It is sensitive to executive function (EF) impairments and has shown consistent results in multiple clinical populations (Ashendorf et al., 2008; Periáñez et al., 2007; Giovagnoli et al., 1996; Lange et al., 2005). Different executive processes are thought to be associated with performance in the TMT, including inhibitory control, working memory, and attention (Arbuthnott & Frank, 2000; Kortte et al., 2002; Salthouse, 2011; Sánchez-Cubillo et al., 2009). In addition to its sensitivity to executive dysfunction, the TMT presents several strengths, as it is simple and intuitive, easy to understand for patients, has a short administration, can be used in different cultures, and the existence of adapted versions allow cross-cultural comparisons (Kim et al., 2014; Lee et al., 2000; Maj et al., 1993).

However, the standard version of TMT presents severe limitations. First, its multiple underlying EFs are not well discriminated, which makes it a test with a low specificity. Solving the TMT involves the articulation of multiple processes (e.g. motor preparation and execution, visual search, visuomotor planning and coordination, working memory, inhibition, among others). The behavioral scores do not disentangle these processes, and the final performance constitutes a rough summatory and undiscriminated assessment (Sánchez-Cubillo et al., 2009). Second, the results on the TMT include a very limited set of measures, consisting most frequently on the total time for completion. Third, it has high variability, given that only one sample (trial) is measured per condition, and that time is measured with low accuracy. Fourth, the spatial configurations of the targets are fixed, and its effects are largely unexplored; thus, it is currently a confounding factor when comparing part A and part B (Fossum et al., 1992; Gaudino et al., 1995). Fifth, the TMT has moderate accuracy for impared neuropsychological performance (Chan et al., 2015). Taken together, these limitations reveal the necessity of new versions of the task unraveling the underlying EFs process, where time and hand trajectories are measured with more precision, and where spatial configurations are controlled.

In recent years, a few studies have attempted to dissect the TMT into smaller subcomponents, in order to scrutinize in more detail which processes are relevant during the task. Digital versions of the TMT that present a more refined measure of time have been developed, some of them measuring hand trajectories (Salthouse & Fristoe, 1995; Woods et al., 2015; Fellows et al., 2017; Dahmen et al., 2017). Even more scarce are digital versions of the TMT with eye-tracking. To our knowledge, only one eye-tracking study in TMT parsed the task in monitoring and planning measuring the interaction between hand and eye movements (Wölwer & Gaebel, 2002). Monitoring occurred when the eye-fixations were close to the hand, whereas planning occurred when the eyes were far from the hand (Wölwer & Gaebel, 2002). Despite the interesting theoretical and methodological contributions, Wölwer and Gaebel measured eye movements with a low resolution and low sampling rate eye-tracking device, which implicated a serious limitation to measure eye-fixations. Moreover, most of the above mentioned limitations still hold for this pionering report.

Here, we aim to tackle most of the TMT limitations by designing a computerized version (cTMT) with several innovations. Our design measures performance in multiple trials, it has a controlled spatial configuration, and measures hand and eye movements with high temporal and spatial resolution. More importantly, this design allowed us to reveal different underlying processes. We parsed the task in three phases: monitoring, planning, and a new phase called exploration. Exploration consisted of eye movements scrutinising the scene before the first-hand movement in a trial, collecting information of the scene before starting the actual task of concatenating visual targets. We also validated the cTMT by comparing its performance with the classical TMT and by testing its association to executive functions assessed by a standard neuropsychological battery. We reported specific EFs underlying the task but also externally validated. To that end, we investigated internal markers of working memory and inhibitory control.

Based on the antecedents and our design, we present a specific set of hypotheses. We anticipate that (a) cTMT will parallel classic TMT outcomes and will be well validated with external measures of EFs. Given its higher complexity, (b) TMT-B will exhibit differential eye-movement features in relation to TMT-A. For instance, TMT-B will present more eye fixations than TMT-A. (c) Some of these eye-movement features will reflect the higher-order EF demands in TMT-B that are not present in TMT-A (d) A subgroup of features will be associated with individual differences in performance. A novel (e) eye and hand marker of working memory and inhibition will be obtained from TMT and will correlate with external EF measures.

Our study showed that measuring hand and eye movements during the TMT can lead to important clinical applications, by revealing different underlying EFs, extracting more refined information, and potentially reflecting particular deficits in clinical populations.

## 3. Methods

### 3.1. Participants

Sixty-one participants were evaluated with the computerized version of the cTMT. Participants reported no record of neurological or psychiatric disorders and no consumption of psychotropic drugs. From this sample some participants had to be excluded from the analysis: 12 participants due to poor data acquisition or not following the instructions. The final sample consisted of 49 participants (24 women, between 18 and 42 years old, mean +/- std = 25.7 +/- 5.4), except for additional online measurements (see below), which consisted of a final sample of 41 participants (16% of attrition rate ~1 year later). This dropout rate does not affect the results, since power analysis showed that from 12 subjects the empirical power was above 0.99 for performance measures (Monte Carlo simulation, 10000 iterations; library *MKpower* in R language (Kohl, 2020) ). All subjects were naïve to the objectives of the experiment and had normal or corrected-to-normal vision. All the experiments described in this paper were reviewed and approved by the IRB of CEMIC Medical Center and qualified by the Department of Health and Human Services (HHS, USA): IRb00001745 - IORG 0001315. All participants provided written informed consent in agreement with the Helsinki declaration.

## 3.2. Computerized TMT (C-TMT)

### 3.2.1. Procedure

The task follows the original design of the TMT (Fig. 1A) (Bowie & Harvey, 2006). Participants had to connect 20 items in consecutive order. In TMT-A, only numbers are presented (1 to 20). In TMT-B, both numbers (1 to 10) and capital letters (A to J) were presented. Participants have to connect items in alternate order, starting from number 1 (1, A, 2, B, and so on). The complete task was divided into 5 blocks of 20 trials, divided by four breaks for resting. Each trial started when participants pressed the left mouse button (Fig. 1B). As soon as they pressed the button, the stimuli appeared on the screen, and they had to pass over every item without releasing the button. When the mouse button was released, the stimuli disappeared, and a fixation dot appeared. Each trial had a time limit of 25 seconds. Trials ended when participants released the button or when they reached the time limit.

Every block started with a drift correction for the eye-tracker in which participants had to fixate in a small circle (20 pixels) and press the spacebar (Fig. 1B). After the drift correction, a small red/blue dot indicated the upcoming trial type (blue and red predicted trials A and B, respectively), and the new trial began with the button press. Participants were instructed to rest between blocks as much as they needed and to resume the task whenever they were ready. Before resuming the experiment, they performed the drift correction, consisting of a central dot in which they have to fixate. If the program failed to detect the eye or if the drift exceeded 2 degrees (EyeLink default value), the experiment stopped and could only be resumed after the participant called the experimenter and a recalibration was launched (built-in Eyelink toolbox function).

Participants completed a total of 100 trials, 50 were TMT-A and 50 were TMT-B, strictly alternating between the two trial types. The task took between 40 and 60 minutes including eye tracker calibration and re-calibrations. The stimuli were presented using Psychophysics Toolbox Version 3 (Brainard, 1997).

### 3.2.2. Stimuli

The spatial distributions of items were the same for all participants, but the order and whether it corresponded to a trial-type A or B was randomized across participants. With regards to the stimuli spatial arrangement design, the item positions were selected one-by-one from a 30-by-30 grid. First, the starting position was selected randomly. Second, horizontal and vertical displacements were selected from a Poisson distribution with the parameter $\mu = 5$. The position was added to the path if the stroke did not cross any previous stroke (straight lines that connected the center of each item, if they were connected in order). After filling the grid with 20 items, the area of the convex hull of the resulting path was calculated. Target arrangements were accepted only if they presented an area larger than 40% of the total area covered by the grid.

Each position of the grid was separated by 20 pixels, which correspond to 0.44 degrees of visual angle. The grid covered 600 x 600 pixels. Each item was a single-digit/character surrounded by a circle with a radius of 10 pixels, centered in a given position of the grid. Finally, several spatial distributions were generated and 100 of them were selected (some examples are presented in Fig. 1C). The final area covered by the convex hull was (50±7)% of the total area covered by the grid.

## 3.3.    Eye-tracking recordings

Participants were seated in front of a 19-inch screen (SyncMaster 997 MB, 1024x768 pixels resolution, 100 Hz refresh rate; Samsung, Suwon, Korea) at a viewing distance of 65 cm, subtending an angle of 29.3 degrees horizontally and 22.5 degrees vertically. A chin rest that was aligned with the center of the screen prevented head movements. An EyeLink 1000 eye-tracker (SR Research Ltd., Ottawa, Ontario, Canada) was used to record gaze locations of both eyes at a sampling rate of 1 kHz. Nominal average accuracy is 0.5 degrees, and spatial resolution is 0.01 degrees root mean squared, as given by the manufacturer. The participant's gaze was calibrated with a standard 9-point grid for both eyes. Built-in drift correction was performed before every block of 20 trials. Based on the results of the drift measures, the participant moved forward in the experiment or had to call the experimenter to perform a new calibration.

The best-calibrated eye was selected for each participant based on the visual exploration of every trial. All eye movements were labeled as fixations, saccades, and blinks by the eye-tracker software using the default thresholds for cognitive experiments (308/s for velocity, 8,0008/s 2 for acceleration, and 0.18 for motion; Cornelissen et al., 2002).

## 3.4.    Data Analysis

### 3.4.1.    Performance Analysis

Each trial had a time limit of 25 seconds for its completion. Given that most of the participants failed to reach 20 items, we decided to use a criterion of 12 correctly concatenated items, starting from the first item, to declare the trial completed and define the response time as the time needed to concatenate the first 12 items. A similar criterion was used for the percentage of completion: the percentage of trials that had been successfully completed until target 12 (PC). It is worth mentioning that increasing the number of items covered throughout the trial significantly reduced the difficulty of the task (even for the first 12 items), due to a benefit in searching the next item among fewer distractors in each step of the task. The selection of the threshold of 12 targets resulted in a good estimation of performance keeping a reasonable amount of data. In fact, the main results did not depend on the threshold (see Fig. S1 for a replication of the results with two other thresholds). Moreover, this criteria generated robust results throughout the task, given that there were no significant learning effects, as revealed by the comparison of the first and last thirds of the trials regarding the ratio (B/A) for PC and RT (Wilcoxon signed-rank test: PC Ratio: p=0.14; z: -1.49; RT Ratio: 0.12 ; z= 1.56).

Correct trials were those that fulfilled the completeness criterion and also presented a correctly concatenated sequence of targets. To that end, the drawn trajectory of the mouse should enter all the targets only in the correct order (e.g. 1-A-2-B, etc). In order to define a path as correctly concatenated, we evaluated the sequence of items produced by the participant. A target was reached when a threshold of 10 pixels from the center was reached. An additional criterion was that trajectories should not cross. In other words, the trajectory curve should not touch itself, as in the original TMT.

### 3.4.2. Statistics in eye data

In order to compare the distributions of saccades and fixations of cTMT-A and cTMT-B (see Fig. 2C,D,E,F), we filtered the raw data by keeping only the correct trials, discarding fixation durations that were over 1000 ms, and removing saccade durations that were over 100 ms. Finally, to equilibrate the samples, we subsample by an order of magnitude. Given the large amount of data, the distributions didn't change visually after filtering. We applied the Kuiper's test (Kuiper, 1960; library *twosamples* in R (Dowd, 2020)) to statistically test for differences between cTMT-A and cTMT-B.

### 3.4.3. Parsing into stages

Following Wölwer & Gaebel (2002), fixations were classified based on their relationship with the mouse position in three different stages. Fixations were defined as *Monitoring* fixations if they were located near the cursor (closer than 25 pixels) at any time during the fixation interval. Fixations were defined as *Planning* fixations if they were located far from the cursor (farther than 25 pxs) during all the fixation intervals (Wölwer et al., 2003, 2012; Wölwer & Gaebel, 2002). From the planning fixations, we also defined a separate group called Exploration fixations that correspond to the fixations occurring before the first-hand movement.

The mean number of fixations and the median fixation duration were calculated for each participant and condition. Wilcoxon signed-rank tests were used to compare between conditions cTMT-A and cTMT-B.

### 3.4.4. Internal measure of Working Memory

A remembered target is one that was seen and not immediately selected with the mouse, i.e., other targets were seen before actually passing with the mouse on top of that target. For instance, in Fig. 4Ai, the target "2" was seen while searching for "1", and then reached with the mouse directly without fixating on it; in Fig. 4Aii, the participant saw the target "2" again right before they selected it with their hand. Thus, in the former case of this schematic example, the participant remembered the position of the target, and in the latter, they did not. This criterion does not differentiate if there is one or several targets between the last view and the passage with the mouse. This analysis only included correct trials (as defined in section 3.4.1).

Regarding the target remembered ratio (TR-B/TR-A) along with the task, we calculated the previously described metric in 5 blocks with 20 trials each. These were the actual blocks of the task, with a pause between them (Fig. S3). The target remembered ratio was calculated for each block separately. For this particular metric, only blocks with at least 3 correct trials for each part were included.

### 3.4.5. Internal measure of Inhibitory Control

Hand trajectories are directed towards the fixated targets (Bulloch et al., 2015; Fisk & Goodale, 1985), when they are the next in the sequence (Correct Detections). Inhibition occurs during fixations on items that do not follow in the sequence (False Detection), when the hand must keep its trajectory without orienting it towards the item (Fig. 5A). A lack of inhibition will be manifested as a persistent tendency to orient the hand trajectory towards False Detections.

In order to add all the hand trajectories projected into the direction of the fixated item, first, the hand trajectories were segmented between the onset and the offset of the fixations into the items. Second, the position of the hand at the time of the fixation onset was subtracted in both vertical and horizontal directions. Therefore, all the trajectories start at the origin (0,0). Third, they were projected into the direction of the fixated item and normalized by the distance between the initial point and the item. Thus, the fixated item was at the point (1,0). Finally, fixations to the next item in the sequence (Correct Detections) were separated from fixations to other items (False Detections) (see Fig. 5A).

The spatial distribution of the trajectories was estimated as the 2D-histogram of the trajectories (see Fig. S4). The temporal pattern was estimated as the position relative to the fixated item (and projected as described before) as a function of time.

## 3.5. External validation measures

In order to perform an external validation of the cTMT measures, we administered an executive functions battery (Torralva et al., 2009) and a canonical visual working memory test, the Change Detection Task (CDT) (Luck & Vogel, 2013). The CDT was implemented online, using the jsPsych library (de Leeuw, 2015) in JavaScript language and deployed in the Cognition platform ([www.cognition.run](www.cognition.run)).

### 3.5.1. INECO Frontal Screening

The INECO Frontal Screening (IFS) was collected as recommended by the validation study (Torralva et al., 2009). The IFS evaluates EFs providing high sensitivity to characterize deficits among different clinical populations (Bahia et al., 2018; Broche-Pérez et al., 2019; Ihnen et al., 2013; Torralva et al., 2009). The IFS includes a Motor Programming task (Dubois et al., 2000); Interference (Dubois et al., 2000) and a Go/NoGo (Dubois et al., 2000) tasks based on motor sequences; a Verbal Inhibitory Control task (Burgess & Shallice, 1997) in which participants have to complete the final word of a sentence, avoiding its strong constraint; a Verbal (Hodges, 2017) and a Spatial (Wechsler, 1987) Working Memory tasks, a Backward Digit Span (Hodges, 2017); and a measure of Abstraction Capacity by reporting proverb interpretations (Hodges, 2017). Each task adds points that sum up to a total between 0 and 30. Using a cutoff of 25 points, sensitivity of the IFS was 96.2%, and specificity 91.5% in differentiating controls from patients and it correlated with classical executive tests such as the time to complete TMT-B (r = −0.75; p < .001) (Torralva et al., 2009). The IFS has good internal consistency ($\alpha$ = 0.80), sensibility to evaluate frontal-executive dysfunction (Moreira et al., 2019), and was remarkably similar with increasing age (Sanjurjo et al., 2019). The IFS was administered through an interview with the experimenter, and overall it took approximately 10 minutes.

### 3.5.2.  Change Detection Task

The change detection task is a simple assessment that can reliably estimate visual working memory capacity (VWM) in a very simple way (Luck & Vogel, 2013). An array of 4 or 6 colored squares were presented for 150 milliseconds and after a 900 milliseconds interval with no stimuli, only one square appeared on the screen. Participants had to respond if that square was part of the original array or not, meaning that it had the same colour as the one presented in the array in that particular position. Subjects responded using two keyboard keys with the index finger of each hand. There were consistent and inconsistent types of trials. In our online experimental design (**Fig. 4D**) we show 120 trials and response times (RT) and keyboard responses were measured. In order to evaluate VWM capacity, we calculate the number of items stored in working memory on a given trial type (K) **(eq. 1)** for the 6 array trials ($K_6$), 4 array trials ($K_4$), and the average value between them ($K_{average}$).

$$K = N_{set}(2\frac{correct\ trials}{all\ trials} - 1) \tag{1}$$

Where $N_{set}$ corresponds to the number of squares in the presented array for a specific trial (4 or 6).

# 4.  Results

## 4.1.  Hand movements: global performance of the cTMT

Participants completed 100 trials of the cTMT, strictly alternating between part A and part B (Fig. 1A, but note that we also replicated the relevant results in a subsample of 30 trials). A trial started when the participant pressed the mouse button, which enabled them to draw on-screen, and finished when the participant released the mouse button or after 25 seconds. We applied this time limit in order to run the whole experiment in approximately 40 minutes, avoiding fatigue (the total time for completing 30 trials is less than 12 minutes). As a consequence, participants didn't have time to reach all the items in many cases. Thus, to characterize the general performance we used the hand movement's data and measured both the time needed to concatenate 12 targets in the correct order (RT) and the percentage of trials that had been successfully completed until target number 12 (percentage completed, PC).

Regarding validation measures using hand movements, the initial mouse button press, and the final button release, we found a significant increase in RT in part B compared to part A (Wilcoxon signed-rank test: $p=1.1*10^{-9}$, $z=-6$; Fig. 1B). Also, the PC was lower in B compared to A (Wilcoxon signed-rank test: $p=1.1*10^{-9}$, $z=6$; Fig. 1C). It is important to note that these results hold even considering only the first 30 trials, including both conditions, and excluding the very first ones to discard possible errors (trial 1 from TMT-A and trial 1 from TMT-B) (see Fig. S2A,B). This is consistent with previous results for the pencil and paper TMT task (Margulis et al., 2018; Tombaugh, 2004).

Next, the performance based on hand movements was tested for associations with EFs. We observed a significant correlation between the Completion Ratio (PC-B/PC-A) and the total IFS score (Fig. 1E; Spearman Correlation: $R=0.437$, $p=1.7*10^{-3}$), but not with the RT ratio (Fig. 1D; Spearman Correlation: $R=0.006$, $p=0.97$). These results served as a validation of the proposed version of the TMT. It is worth mentioning that the fact that the correlation of the IFS with PC, but not with the RT was significant, might be a direct consequence of the time pressure and time limit of our design, not present in the original TMT.

One relevant aspect of the present version of the TMT is that spatial configurations are extensively explored and, crucially, they were the same for all participants, except that trial type (A or B) was assigned randomly to each spatial configuration before the experiment. As seen in Fig. 1F,G, the initial hand trajectory is similar in both parts, rich in twists in order to not overlap the trail. As the vast majority of participants couldn't reach the last targets, It is common to find at the end of the B trials a decrease in the density of samples (see Fig. 1G).

In summary, the novel cTMT results resemble the classic TMT, even taking only the first 30 trials. These results are exclusively related to the change of lists (only numbers or numbers and letters) as the same spatial configurations were presented in both types of trials. Finally, the performance correlated with an external screening test of EFs (IFS).

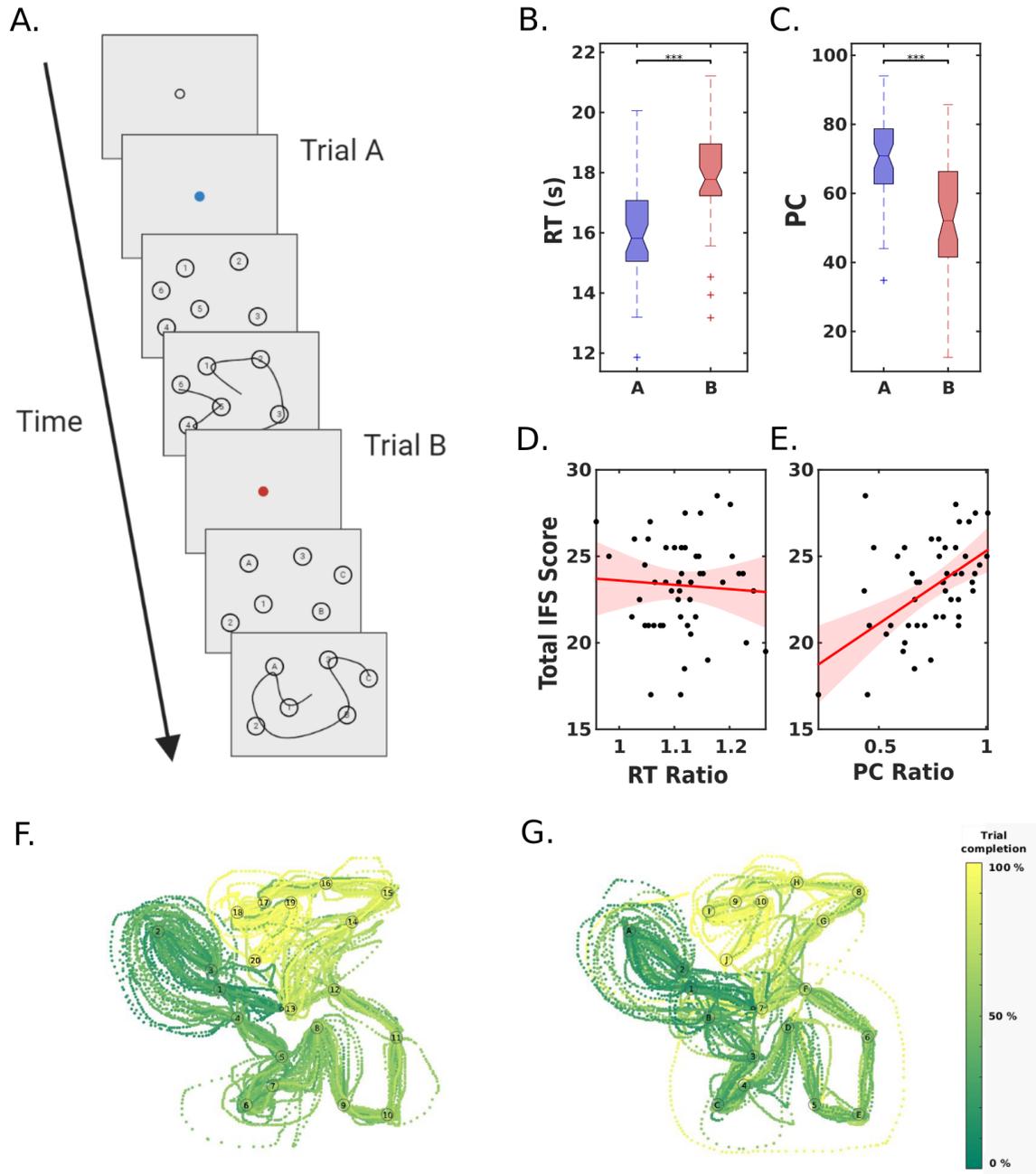

**Figure 1: A.** Experimental design and task validation. The trial begins with a mouse button press and continues until the mouse button is released or a maximum time of 25 seconds is reached. **B.** Time to connect 12 items in order (completion criteria), and **C.** Percentage of completed trials (PC) for both Part-A and Part-B. **D.** Correlation between the Total IFS Score and the Completion Ratio (PC-B/PC-A). **E.** Correlation between the Total IFS Score and the RT Ratio (RT-B/RT-A) **F.** Hand trajectory in an A type trial. **G.** Hand trajectory in a B type trial, with the same configuration. The colorbar represents the relative temporal evolution for each subject.

## 4.2. Eye movements

We observe a similar structure in both scanpaths, except that TMT-A has, qualitatively, more color consistency along the trajectory, revealing that in TMT-A almost all trials reached the last item. Fig. 2A,B illustrates the eye scanpaths of two representative trials (TMT-A and TMT-B, respectively) with identical spatial configuration. TMT-B, on the other hand, presented more variability in the number of reached items, also reflected by the larger error bars observed in the PC barplot in Fig. 1C.

Regarding fixations and saccades, TMT-A and TMT-B were indistinguishable in many measures, including saccade and fixation duration as well as saccade amplitude (Fig. 2C,D,E; Kuiper's test: V=0.01, p=0.77; V=0.01, p=0.96; V=0.02, p=0.11 respectively). The number of fixations showed a clear difference between both conditions, with a higher number in TMT-B (Kuiper's test: V=0.17, p = $2.5*10^{-4}$; Fig 2F). An identical saccade duration, saccade amplitude and fixation duration suggests a similar visual mechanism between parts A and B. The observed difference in the number of fixations might originate in a more complex processing of the task, more related to higher-order cognitive processes than visual mechanisms. In other words, to solve both parts of the task, subjects seem to use their visual machinery in a very similar way, except that TMT-B requires a more intensive scanning of the visual scene. To explore in more detail the possible mechanisms involved in the differential performance of A and B, we parsed the task in three phases, and analyzed eye movements in each phase.

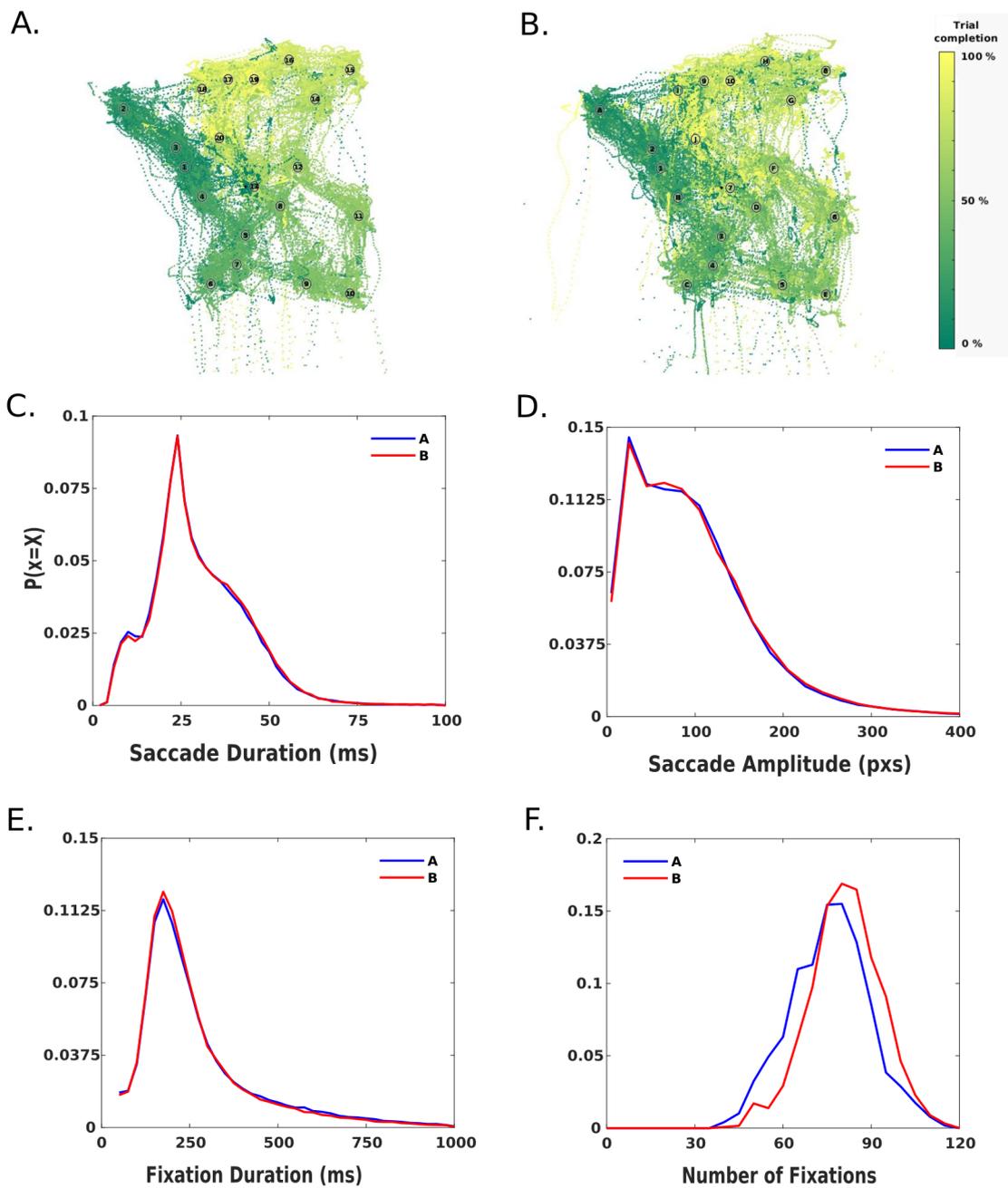

**Figure 2: A,B.** Eye trajectory for all the subjects in a TMT-A (**A**) and TMT-B (**B**) trials. The colorbar represents the relative temporal evolution for each subject. Note that the yellow colour, representing the final fixations of the trial, are more consistently located around the last targets. **C,D,E,F.** Individual eye movements characteristics. Distributions of saccade durations (**C**), saccade amplitude (**D**), fixation durations (**E**), and the number of fixations during the trial (**F**).

## 4.3. Parsing the task in three phases using hand and eye movements

The previous section showed that the difference in the time needed to complete the task in both conditions is mainly explained by the number of fixations performed during the trial, and not by fixation duration or saccade duration. In the following section, we will focus on the number of fixations in our analysis. We aimed to understand which aspects of the resolution of the task change between parts A and B, revealed by fixation type.

Previous work classified fixations during the TMT in two phases: planning and monitoring (Wölwer & Gaebel, 2002). Here, we use a similar classification, with the addition of a new initial exploration phase. It corresponds to all fixations occurring at the start of each trial before the movement of the hand and accounts not only for the search of the first item but also for the initial exploration of the scene (Fig. 3A). The monitoring phase consisted of fixations that occurred over the cursor and were more related to the motor execution of the task, while the planning phase consisted of fixations that occurred outside the cursor and were related to more executive aspects of the task. For an illustration of the phase classification, we created a video where fixations are colored according to the phase where they occur in real-time (See Supplemental Material).

First, we compared the number of fixations and fixation duration between TMT-A and TMT-B at each phase. We observed a higher number of exploratory (Wilcoxon signed-rank test: $p=9*10^{-4}$, $z=-3.3$) and planning fixations in part B (Wilcoxon signed-rank test: $p=1.9*10^{-9}$, $z=-6$), following the trend of the overall task. Conversely, there was a lower number of monitoring fixations in part B (Wilcoxon signed-rank test: $p=1.1*10^{-4}$, $z=3.8$; Fig. 3B). Regarding fixation duration, there was no significant difference between TMT-A and TMT-B in exploration (Wilcoxon signed-rank test: $p=0.98$, $z=-0.01$) and planning (Wilcoxon signed-rank test: $p=0.7$, $z=0.38$). However, in the monitoring phase, fixation duration was higher in TMT-A (wilcoxon signed-rank test: $p=2.4*10^{-6}$, $z=4.7$; Fig. 3D). Fig. 3B,D shows that the number of fixations was more informative than fixation duration, explaining the differences between TMT-B and TMT-A, which is consistent with the distribution of eye movements depicted in Fig. 2.

To explore the association between eye movements and performance in the three phases of the task, we calculated the ratio (B/A) of the number of fixations and their corresponding duration , and correlated them with a measure of performance (RT-B/RT-A). We found significant correlations between RT ratio and the number of fixations ratio in exploration (Spearman Correlation: $R=0.39$, $p=5.2*10^{-3}$) and planning phases (Spearman Correlation: $R=0.29$, $p=4.3*10^{-2}$; Fig. 3B,C), but not in monitoring (Spearman Correlation: $R=-0.22$, $p=0.13$). This is again consistent with the differences in the number of fixation's distribution (Fig. 2F). As seen in the distributions in Fig. 2E, fixations' duration didn't vary between conditions, so it was expected that the fixations' duration ratio didn't affect the RT ratio (Spearman Correlation in Exploratory: $R=-0.05$, $p=0.71$ ; Planning: $R=-0.02$, $p=0.91$ ; Monitoring: $R=-0.05$, $p=0.74$; Fig. 3E).

Summarizing, fixation number but not fixation or saccade duration/amplitude varied between parts A and B and provided adequate measures of task performance. Splitting the task in three phases unveiled the different aspects of the executive process (exploration, planning, execution and monitoring). The increase in the number of fixations in B versus A, in both exploration and planning; as well as a decrease in monitoring characterized the different stages. Additionally, a small but significant increase in fixation time in part A versus B was observed only in the monitoring phase. Lastly, the increase in the number of fixations observed in B/A for the exploration and planning phases correlated positively with relative performance B/A (RT Ratio).

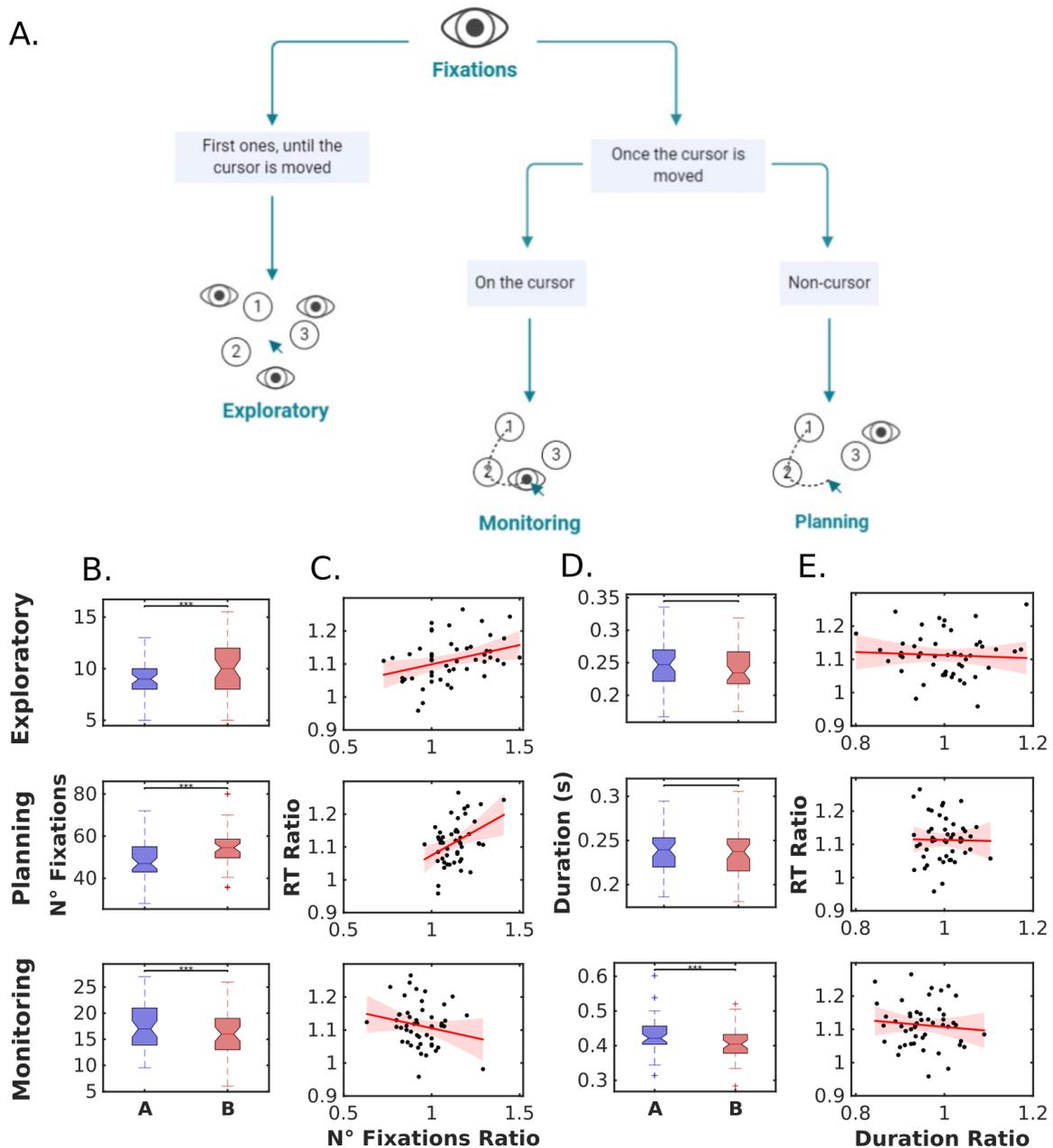

**Figure 3: A.** Phases of the TMT-Task, in terms of hand-eye interactions. This diagram classifies fixations according to the three phases where they occured: exploratory,

monitoring and planning. **B.** Boxplots of the median number of fixations in each of the three phases (exploratory, planning and monitoring) for both parts (TMT-A and TMT-B). **C.** Correlations between the RT Ratio (RT-B/RT-A) and the number of fixations' ratio in each phase. **D.** Boxplots of the median duration of fixations in each phase (exploratory, planning and monitoring) for both parts (TMT-A and TMT-B). **E.** Correlations between the RT Ratio (RT-B/RT-A) and the duration of fixations ratio in each phase.

## 4.4.  Visual Working Memory

In this section, we derived an internal measure of visual working memory using eye and hand movements in the cTMT. Then, we inspected how this internal measure of visual working memory of the targets affected performance in TMT-B with respect to TMT-A. Finally, we compared the derived measure with the individual performance in a validated visual working memory task. In order to quantify our measure, we estimated the number of Targets Remembered along the search (TR), i.e. the number of targets which had no fixations right before the hand reached them (see methods section), including only correct trials.

On average, participants remembered more targets in TMT-A than in TMT-B (Fig. 4B; A=4.60±0.87; B=4.38±0.98; Wilcoxon signed-rank test: p=0.028; z=2.2). This result is consistent with fewer overall fixations in TMT-A given that a higher target location memory implies less target search around the scene (see Fig. 2). It is also consistent with fewer planning fixations (see Fig. 3). This suggests that participants memorized the location of more targets ahead and had to look again at the same target fewer times in order to correctly complete the trial in TMT-A.

The TR Ratio (TR in B / TR in A) correlated with the PC Ratio (Fig. 4C; Spearman Correlation: R=0.48, p=5.3*10$^{-4}$), indicating that the relative improvement in remembering targets in B was associated with the overall performance of the task. Moreover, the TR ratio was tested for associations with an external WM measure, the visual working memory capacity (K$_{average}$) estimated from a Change Detection Task (Fig. 4D) (see Methods) (Luck & Vogel, 2013), showing a strong correlation (Fig. 4E; Spearman Correlation: R=0.43, p=5.3*10$^{-3}$, N=41).

In brief, we extracted a novel internal measure of visual working memory in the cTMT that correlated both with performance (PC Ratio) and a canonical external VWM measure (CDT), suggesting that it is possible to isolate individual EF components within the cTMT.

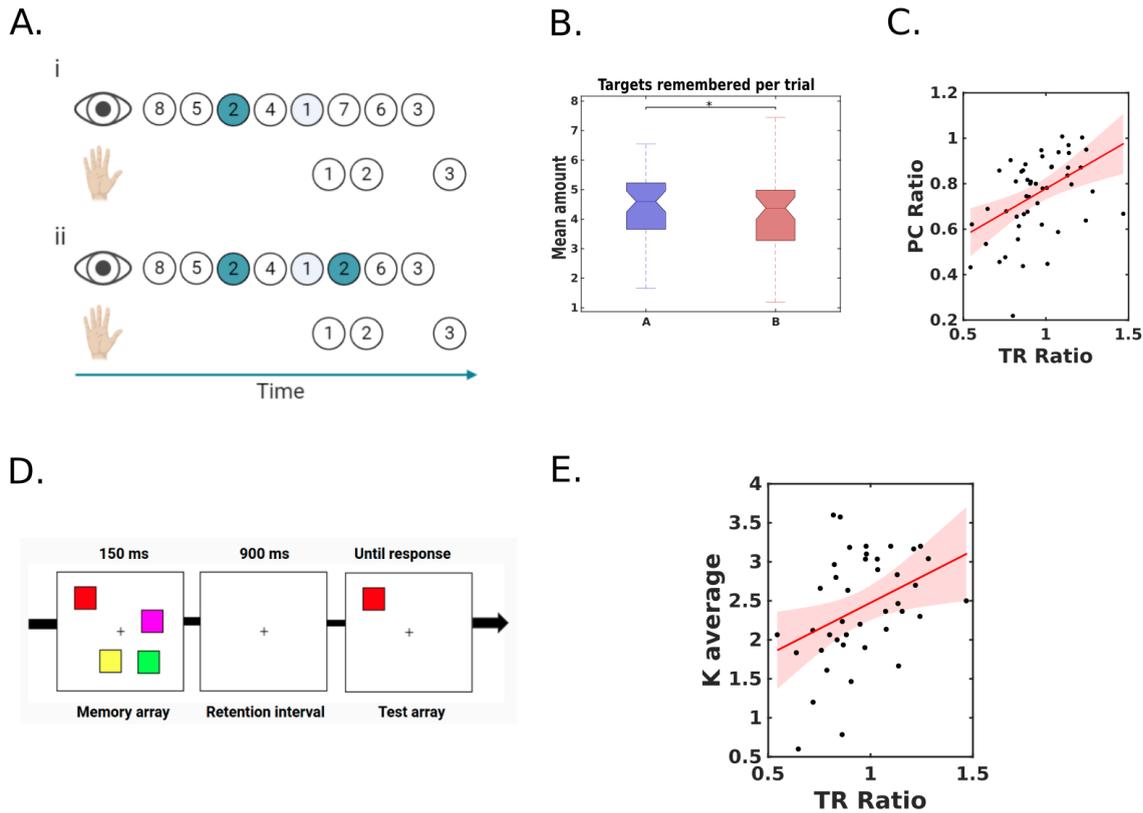

**Figure 4: A.** Diagram for identifying targets remembered. In **4.A.i** the target "2" is remembered as it wasn't looked right before reaching it with the cursor, while in **4.A.ii** the target "2" wasn't remembered as it was seen again right before reaching it with the cursor **B.** Boxplots for the mean amount of Targets Remembered (TR) for each subject in both parts (TMT-A and TMT-B) **C.** Correlation between the PC Ratio (PC-B/PC-A) and the TR Ratio (Targets Remembered in B/ Targets Remembered in A). **D.** Experimental design of the Change Detection Task. Memory array: 4 or 6 colored squares was shown on screen during 150 ms, retention interval: only fixation cross through 900 ms, Test array: A single square appeared on screen with same color and location for consistent types of trials and difference in one or both characteristics for inconsistent ones **E.** Correlation between Visual Working Memory Capacity ($K_{average}$) and the TR Ratio.

## 4.5.    Inhibitory control

In this section we derived, from the eye and hand movements' data, a second internal measure of executive functioning, in this case inhibitory control. When the eyes fixate on a new item, it could be either the next item in the sequence or not, i.e. it could be a Correct or a False Detection of the target. In the latter case, the hand has to avoid following the eye and wait until the correct item is found. This behavior is evident when aligning all the paths explored by the hand after fixating a new item (Fig. 5B,C, S3). The spatial distribution of these paths shows that, when a correct item was identified, the hand moved directly towards the target (Fig. 5B). When a false detection occurred, the hand stayed still (Fig. 5C) or moved in other directions (Fig. S4) showing an inhibition of early motor actions. In order to quantify this behavior, we estimated the displacement in the direction of the new item, for Correct and False detections, and for TMT-A and TMT-B separately.

Consistent with the spatial distributions, the hand displacement for the Correct detections was larger than for False detections, reaching the position of the target (displacement=1, Fig. 5D) and revealing an inhibitory motor process. Interestingly, the curves in TMT-A and TMT-B were similar, as it was also evident for the difference curves between Correct and False detections (Fig. 5E). The Area under the difference curves was significantly different from zero for both cTMT-A and cTMT-B (Fig. 5F; Signed Rank test: TMT-A: $p < 10^{-8}$, TMT-B: $p < 10^{-8}$), but there was no significant difference between TMT-A and TMT-B (Signed Rank test, p=0.29).

The Area under the difference curves could be an interesting estimation of the inhibition, the larger the area, the larger the inhibition to avoid following the eyes after False detections. Nevertheless, the Displacement in both TMT-A and TMT-B did not yield significant correlations with the IFS (Spearman correlation, rho<0.2, p>0.25, N=49) or its subset of verbal inhibitory measures (Spearman correlation, rho<0.15, p>0.45, N=49).

In summary, we extracted a novel internal measure of inhibition in the cTMT that seemed to capture the dynamics of inhibitory control processes within the task, but it did not reflect the difference in performance, and it did not correlate with the external measures (IFS).

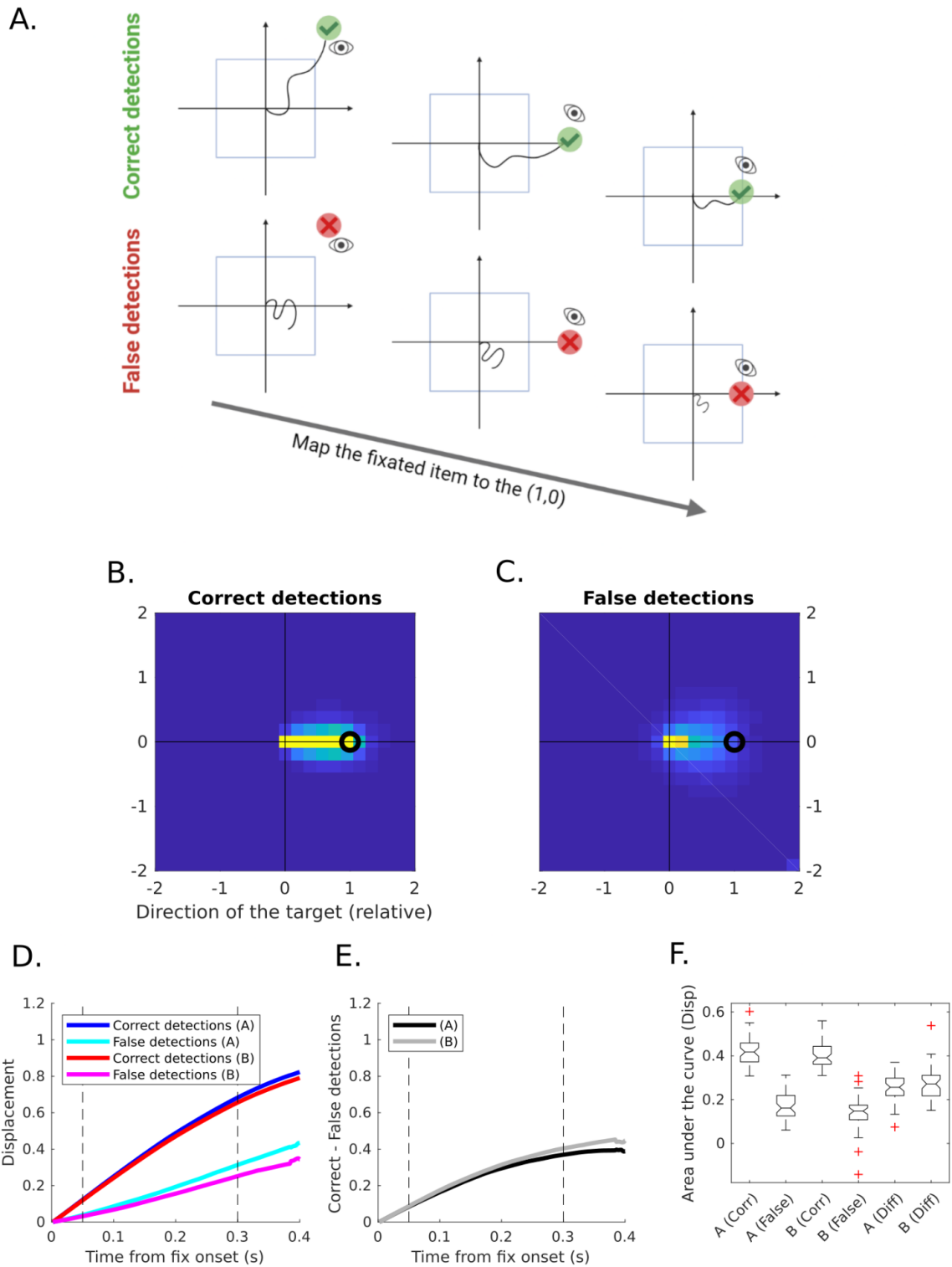

**Figure 5: A.** Diagram showing the difference between the Correct detections and the False detections. **B,C.** Spatial distribution of the paths explored by the hand when fixating a new item, for Correct (next in the sequence; B) and False (C) detections. 2-D paths were aligned and normalized so that the fixated item was at (0,1) (see Fig. S4). **D.** Relative displacement towards the fixated item, in the direction of the item. Curves are aligned to the fixation

onset. Red: TMT-A, Correct; Magenta: TMT-A, False; Blue: TMT-B, Correct; Cyan: TMT-B, False. **E.** Difference between Correct and False detections for displacement. Black: TMT-A; Grey: TMT-B. **F.** Area under the difference curves for the displacement. The area was calculated participant-by-participant between 50 and 300 ms.

## 5.  Discussion

In the present study, we aimed to design a computerized version of the TMT (cTMT) that could tackle its main limitations. In particular, we aimed to build new measures within the task that could reflect individual EF processes, based on the precise recording of hand and eye movements. Firstly, we validated the cTMT showing that the RTs and performance profiles are consistent with the classic TMT. Moreover, we observed a significant correlation between the Completion Ratio and an independent executive functions battery (IFS). Secondly, we showed that eye movements' features were very similar in TMT-A and TMT-B, and differed only in the number of fixations, implying that the visual mechanisms are similar between conditions but they differ in more higher level processes. Thirdly, when the task was parsed into three different stages (exploratory, planning and monitoring), we found a higher number of exploratory and planning fixations in TMT-B, and a lower number of monitoring fixations. This could be interpreted as higher planning and executive (high level) costs in B, and less resources devoted to lower level processes (monitoring hand movements). Fourthly, the mean amount of targets remembered was higher in TMT-A, and the ratio of remembered targets between TMT-B and TMT-A correlates with the Completion Ratio of the whole task. These results imply a lower memory performance in part B given its higher demands, and that the individual memory skills within the task explain, at least in part, overall performance. Strikingly, the amount of remembered targets also correlated with an external measure of visual working memory capacity ($K_{average}$ in Change Detection Task), which validated our measure as an individual marker of VWM. Finally, we derived a potential internal measure of inhibition that is based on the hand movements towards Correct and False eye detections of items. To our knowledge, this is the first study that uses high resolution eye and hand movements in TMT. One important aspect of our work is that we were able to dissect the task and extract individual markers of EFs, tackling one of the main limitations of the traditional TMT, making the cTMT a promising tool for research and clinical use.

As our first hypothesis, we replicated the general results of classical TMT: the resolution of type B trials took more time, while the percentage of completion was higher for A type trials. Furthermore, we found a correlation between the Total IFS Score and the Completion Ratio (PC-B/PC-A), while not with the RT Ratio, probably due to the limiting time factor. Previous digital implementations of the TMT expanded the analysis of the classic version by extracting more features (Dahmen et al., 2017; Fellows et al., 2017; Salthouse & Fristoe, 1995; Woods et al., 2015) but in this work we also focused on extracting internal measures as markers of Executive Functions (EFs).

There were only a few previous experiments on eye movements with the TMT task. One used a high-resolution eye tracker but did not extract any more features other than the number of fixations (Hicks et al., 2013), and others tried to disentangle the task but used low-resolution eye trackers (Jyotsna et al., 2020; Wölwer et al., 2003, 2012; Wölwer & Gaebel, 2002). Thus, we started inspecting eye movements recorded with a high-resolution eye-tracker that enables fixation and saccade analysis. We observed that even though almost all fixation and saccade properties were very similar between both TMT parts (saccade and fixation duration, saccade amplitude), the number of fixations was statistically higher in part B. This result is consistent with previous bibliography (Hicks et al., 2013) and may be the result of increased cognitive load interfering with the participants' search strategy. The number of fixations can assess participants' attention by indicating how many attentional resources are utilized between stimuli (Hyönä, 2010).

From the previous work on eye movements in the TMT, a series of works by Wölwer and colleagues (2003; 2012) and Wölwer & Gaebel (2002) proposed that the task could be divided into phases, and that the total time spent in each phase change in different patient populations. Starting from the taxonomy previously proposed by Wölwer & Gaebel (2002) (monitoring and planning fixations) and adding a new class called exploratory fixations, we explored separately the number of fixations and the fixation duration. Previous work was done using a low-frequency eye tracker (50Hz), and thus their analysis was limited to total time on each phase (Wölwer & Gaebel, 2002). When we focused on the number of fixations, we found a higher number of exploratory (the first ones, until the cursor moves) and planning (those fixations away from the cursor after the first movement) in part B. This is also consistent with Wölwer & Gaebel (2002), who showed that the longer planning periods in schizophrenia patients resulted from a higher number of fixations within such a planning period in both test versions. In relation to the fixations' duration, we only saw statistical differences in the monitoring ones (those after the first movement and over the cursor). This is consistent with the fact that the monitoring phase is more related to the motor execution of the task, but the planning phase is related to more executive aspects of the task (in other words, to the specific executive component needed in TMT-B (Gouveia et al., 2007)). We suggest there's an amount of time participants dedicate to monitor the cursor (without limitation of time practically) in part A. But in B, as it is more complex and more cognitive load is involved, subjects sacrifice this time in order to dedicate it to planning (trade off). Saccade durations are not related to processing costs, and they have a small impact in total time as they are smaller than the fixation durations, and didn't change between part A and B.

Then, we investigated the visual working memory performance based on remembered TMT items. We found a higher number of remembered targets in part A that is consistent with less planning fixations in A, since participants might use their memory of target locations, requiring less search in TMT-A. The TR Ratio correlated with the overall performance (PC Ratio) and also correlated with an external measure of visual working memory (CDT) (Luck & Vogel, 2013), validating our cTMT memory measure. It is worth noting that previous work linked the TMT performance with working memory, but results depended of which tests were administered (canonical and complex tests as the Wechsler Memory Scale and the Wisconsin Card Sorting Test, among others) (Crowe, 1998 ; Larrabee & Curtiss, 1995 ; Mahurin et al., 2006; Kortte et al., 2002; Sánchez-Cubillo et al., 2009). These works focused on correlating results of classical tests in a general way and, to our knowledge, no other reports have attempted to examine the relationship between internal markers of the TMT and specific VWM tests as the CDT.

Based on the cTMT we not only extracted a working memory measure but also a way to assess cognitive inhibition using only the hand and eye trajectories. As Sánchez-Cubillo and collaborators (2009) remarks, the role of inhibitory control (IC) in TMT is not fully elucidated. In accordance to Arbuthnott & Frank (2000), a relationship between TMT-B and inhibitory abilities has been supported on the basis of significant correlations between TMT and the Stroop Interference condition (Chaytor et al., 2006; Spikman et al., 2001). However, the use of more specific measures of inhibitory abilities such as Go/No-Go tasks (Langenecker et al., 2007) or negative priming tasks (Miner & Ferraro, 1998) has provided contradictory evidence about the role of inhibition in TMT scores with both positive and negative results, respectively. In fact, previous work highlights the complexity of this particular executive function, as it represents a multidimensional construct (Sperl et al., 2021) difficult to disentangle (Meyer & Bucci, 2016). In this research, we aimed to use more precise measurements of hand and eye movements within the task to build specific IC estimates. Here, our estimate utilises the Correct and False detections of the next target in order to quantify the inhibitory control of the subjects. Consistent with the spatial distributions, the hand displacement for the Correct detections was larger than for False detections, reaching the position of the target. The Area under the difference curves was significantly different from zero for both types of trial, but not between them. We suggest that the area under the difference curves could be an interesting estimation of the inhibition -the larger the area, the larger the inhibition to avoid following the eyes after False detections-. This was corroborated with our data. Nevertheless, we suffered from the same deficit as previous work, failing to find significant correlations with the external measures of IC drawn from the IFS questionnaire. Future work should explore more deeply other IC tasks such as the Go/NoGo or the Stop-Signal task to disentangle the different aspects of IC expressed in the TMT.

Moreover, these measures were evaluated only on the individual differences in EF in a neurotypical population, which is a demanding test due to the lower intersubject variability. Nevertheless, we highlight the importance of finding correlations between our global performance measures and an independent EF questionnaire, and our WM measure and an external measure of VWM capacity. These results encourage further research to expand the sample to other populations such as FTD / AD where the paper-and-pen TMT has already proven to be very useful, and also previous work found effects analysing the task segmentation (Wölwer et al., 2003, 2012; Wölwer & Gaebel, 2002). The length of the task could be an impediment to evaluate those clinical populations, but in this work we showed that even in this neurotypical population the effects are significant using only in the first trials.

It is worth mentioning that our computerized version of the TMT not only allowed us to record hand and eye movements precisely, but also to overcome some of the limitations of the paper-and-pencil version. For instance, our version balances the spatial configurations for type A and B trials, as the spatial configuration of the targets in the classic version are not the same, implying that part of the results observed might be explained by the particular configurations of TMT-A and TMT-B (Fossum et al., 1992; Gaudino et al., 1995). In fact,

Gaudino and colleagues showed that using only numbers, significant time differences arose between the spatial configurations of parts A and B (Gaudino et al., 1995). So, controlling the spatial configuration allowed us to reduce the sources of variability in time between both parts. Moreover, it lets us make a more accurate conclusion about the task switching, linking these differences in performance with the change of lists: from only numbers to letters and numbers. Additionally, our experimental design had a higher number of trials than previous works (Fellows et al., 2017; Poreh et al., 2012; Salthouse & Fristoe, 1995; Wölwer et al., 2003, 2012; Wölwer & Gaebel, 2002; Woods et al., 2015). But, it is worth noting that there were no significant learning effects, as revealed by the comparison of the first and last thirds of the trials regarding the ratio (B/A) for PC and RT (Wilcoxon signed rank test: PC Ratio: p=0.14; z: -1.49; RT Ratio: 0.12 ; z= 1.56).

## 6.   Conclusions

In recent years computational psychiatry and digital neuropsychology (Germine et al., 2019; Montague et al., 2012) has gained traction based on the use of computational approaches to model neuroscience and behavior variables of interest, and machine learning approaches to predict brain pathologies and syndromes from behavioral measures. One of the limiting factors for using this last type of methods to obtain new insights and develop new tools, is the lack of precise enough measures for executive functions and the extension and diversity of actual protocols. In this way, our contribution could help generate new precise features of different EF based on a single complex task. And, in the future, this task could be even replaced by natural behavior.

The measures presented here will also allow us to understand the internal dynamics and interplay of EF during the resolution of a complex task. To summarize, in the present work, we validated the overall performance of the computerized version of the task with external measures and explored the involvement of eye movements in the different phases of the task resolution in both trial types. Moreover, the cTMT surpasses many of the gaps of the standard TMT: 1) it provides multiple fine-grained subscores of the underlying EFs, which are critical for analyze more specific deficits in different pathologies: 2) this version provides multiple behavioral measures that allow a more robust characterization of the participant's performance and brings multiple feature for machine learning multimodal and multi-feature analysis; 3) it provides greater control of spatial configuration bias and more robust results (less variable) by controlling the potential bias of one single configuration. Thus, we proposed that the cTMT could become a powerful tool for an improvement in the accuracy of diagnoses of a wide variety of pathologies where the EF are affected, such as Alzheimer Diseases or Fronto-Temporal Dementias.

## 7.   Code and Data Availability Statement

The analysis code and the data used in the present study will be available upon publication.

## 8.  Acknowledgements

We thank Sol Fitipaldi and Lucas Sedeño for training IL on the administration and interpretation of the INECO Frontal Screening to the participants. The authors were supported by the Consejo Nacional de Investigaciones Científicas y Técnicas (CONICET) and the Universidad de Buenos Aires (UBA). The research was supported by the Agencia Nacional de Promoción Científica y Tecnológica (PICT 2018-2699) and the CONICET (PIP 11220150100787CO). AI is supported by by grants from CONICET; ANID/FONDECYT Regular (1170010); FONCYT-PICT 2017-1820; ANID/FONDAP/15150012; Takeda CW2680521; Sistema General de Regalías (BPIN2018000100059), Universidad del Valle (CI 5316); Alzheimer's Association GBHI ALZ UK-20-639295; and the MULTI-PARTNER CONSORTIUM TO EXPAND DEMENTIA RESEARCH IN LATIN AMERICA [ReDLat, supported by National Institutes of Health, National Institutes of Aging (R01 AG057234), Alzheimer's Association (SG-20-725707), Rainwater Charitable foundation - Tau Consortium, and Global Brain Health Institute)]. The contents of this publication are solely the responsibility of the authors and do not represent the official views of these Institutions.